# High-Power Near-Concentric Fabry-Perot Cavity for Phase Contrast Electron Microscopy

Carter Turnbaugh[1,3], Jeremy J. Axelrod[1,3], Sara L. Campbell[1,3*], Jeske Y. Dioquino[1,3], Petar N. Petrov[1], Jonathan Remis[4], Robert M. Glaeser[2,3], Holger Müller[1,3†]

Zanlin Yu[5], Yifan Cheng[5,6]

Osip Schwartz[7]

[1]Department of Physics, 366 Physics MS 7300, University of California-Berkeley, Berkeley, CA 94720, USA.
[2]Department of Molecular and Cell Biology, 363b Donner Lab, University of California-Berkeley, Berkeley, CA 94720, USA.
[3]Lawrence Berkeley National Laboratory, One Cyclotron Road, Berkeley, CA 94720, USA.
[4]California Institute for Quantitative Biosciences (QB3), University of California, Berkeley, California 94720, USA
[5]Department of Biochemistry and Biophysics, University of California San Francisco, San Francisco, CA 94158, USA.
[6]Howard Hughes Medical Institute and Department of Biochemistry and Biophysics, University of California, San Francisco, CA94158, USA
[7]Department of Physics of Complex Systems, Weizmann Institute of Science, Rehovot 7610001, Israel
*Present address: Honeywell, Broomfield, CO, USA
†hm@berkeley.edu

**Transmission electron microscopy (TEM) of vitrified biological macromolecules (cryo-EM) is limited by the weak phase contrast signal that is available from such samples. Using a phase plate would thus substantially improve the signal-to-noise ratio. We have previously demonstrated the use of a high-power Fabry-Perot cavity as a phase plate for TEM. We now report improvements to our laser cavity that allow us to achieve record continuous-wave intensities of over 450 GW/cm$^2$, sufficient to produce the optimal 90° phase shift for 300 keV electrons. In addition, we have performed the first cryo-EM reconstruction using a laser phase plate, demonstrating that the stability of this laser phase plate is sufficient for use during standard cryo-EM data collection.**

## INTRODUCTION

Transmission electron microscopy (TEM) of vitrified biological macromolecules (cryo-EM) has become a major source of structural information for molecular biology[1,2,3]. However, cryo-EM is limited by the small phase shift imparted to the electron beam by unstained specimens. In addition, biological specimens tolerate only a limited electron exposure, preventing an improvement to the signal-to-noise ratio (SNR) by increasing the exposure. In particular, the SNR for small proteins remains insufficient to produce 3D reconstructions at atomic resolutions. The smallest currently accessible particles are 50-100 kilodalton (kDa)[4], for example streptavidin (50 kDa)[5,6], hemoglobin (64 kDa)[7,8], and alcohol dehydrogenase (82 kDa)[8]. The low signal-to-noise ratio also increases the number of particles required for a reconstruction.

Biological specimens behave as weak phase objects, applying a small phase shift $\psi(x, y)$ (where x, y are coordinates at the specimen) to the electron wave function, but not absorbing electrons significantly. The intensity of an aberration-free, in-focus image is therefore $\left|e^{i\psi}\right|^2 = 1$. This is independent of the specimen, so the contrast is zero. The conventional technique of acquiring images with a deliberate defocus produces some contrast, but not at low spatial frequencies.

Maximum SNR can be achieved with Zernike phase contrast[9,10,11]. This technique uses a phase plate in a back focal plane to apply a 90° phase shift to the undiffracted portion of the electron wave function, changing the image intensity from $\left|e^{i\psi}\right|^2 = |1 + i\psi + ...|^2$ to $|i + i\psi + ...|^2 \approx 1 + 2\psi$, creating contrast at first order in $\psi$. The Volta phase plate has achieved some success, but suffers from a continuously changing phase shift[12] as well as undesirable loss of signal as electrons pass through the phase plate[13].

Here, we describe progress in the development of a laser-based phase plate for transmission electron microscopy. The laser phase plate—having no materials in the electron beam—can withstand indefinite electron exposure with no charging or incoherent scattering. For compatibility with an existing high-resolution TEM, we designed the phase plate to use a continuous wave laser. To achieve a 90° phase shift, an intensity of hundreds of GW/cm$^2$ is required, so we exploit the resonant buildup of a Fabry-Perot cavity. Phase contrast results from phase shifting the undiffracted electron beam *relative* to the diffracted beam in the back focal plane. Within the finite width of the laser beam, the scattered electrons are phase-shifted commensurately with the electrons that did not diffract at all, so there is no phase contrast for specimen content at the lowest spatial frequencies. The cut-on frequency at which the phase contrast first becomes significant is the lowest spatial frequency beyond the laser beam waist. Thus, to decrease the cut-on frequency, we wish to decrease the mode waist of the beam.

The simultaneous necessity for high-power and small mode waist is atypical for a continuous wave laser cavity. The high-power requirements for a continuous wave cavity have been met, notably in gravitational wave detectors[14]. Cavity high-harmonic generation has reached higher instantaneous intensities, but only with pulsed laser systems[15,16]. The desired mode waists have been realized with continuous wave laser cavities, but at only 1% of the power[17,18]. Thus, our cavity requires especial characteristics to meet our requirements for high circulating power and narrow mode waist.

Previously, we demonstrated the feasibility of a laser phase plate capable of producing a 45° phase shift for electrons at 80 kV (an intensity of 43 GW/cm$^2$)[19]. We now describe subsequent improvements to the cavity mirrors and frequency locking techniques that enable an intensity of over 450 GW/cm$^2$ in the new laser phase plate, providing a 90° phase shift for electrons at 300 kV. Furthermore, we report that the laser phase plate is sufficiently stable to be used for routine data collection with biological samples, as demonstrated by achieving a 3D reconstruction of the 20S proteasome, a 750 kDa protein complex, at a resolution of 3.8A.

## Cavity Design

### Cavity Requirements

The ponderomotive potential describes the interaction between electrons and an electromagnetic wave. As the electron traverses the potential, it acquires a phase shift[20]. For horizontally polarized light in the fundamental mode of a Fabry-Perot cavity, the phase shift near the cavity mode waist is given by

$$e^{2y^2/{w_0}^2}\sqrt{\frac{8}{\pi^3}}\frac{\alpha}{\beta\gamma}\frac{P{\lambda_L}^2}{mc^3 w_0}\frac{1}{2}\left[1+cos\left(4\pi x/\lambda_L\right)\right],$$

where *x* and *y* give the position of the electron beam relative to the laser beam, $w_0$ is the beam waist at the focus, $\lambda_L$ is the laser wavelength, *P* is the one-way circulating power, *m* is the electron mass, $\alpha$ is the fine structure constant, $\beta$ is the electron speed in units of the speed of light *c*, and $\gamma = \frac{1}{\sqrt{1-\beta^2}}$ [20]. We use 1064nm light due to extensive optical component manufacturability and high-power source availability. Therefore, to achieve a 90° phase shift for electrons at 300 kV, we must have $\frac{P}{w_0} \approx 155$ MW/cm. To reduce the waist, and thus improve both the cut-on frequency of the phase plate and the maximum phase shift for a given circulating power, we operate the cavity near concentricity. The combination of 97 kW circulating power and 8.5 µm mode waist (a maximum standing-wave intensity at the focus of 342 GW/cm$^2$) yields the desired phase shift at the focus of the beam. In practice, a higher circulating power is used to compensate for imperfect alignment to the undiffracted electron beam (i.e. non-zero x and y in equation 1).

**Cavity Mechanics**

In addition to the optical requirements, our cavity—with its support structure and actuators—must be small enough to fit inside the electron microscope: the cavity diameter must be less than 25 mm, and the cavity length must be less than approximately 50 mm. To satisfy these demands and provide the necessary degrees of freedom for supporting the laser mode, we use a custom-built monolithic aluminum cavity mount (UC Berkeley Physics Machine Shop) with an integrated flexure between the two mirrors. To avoid perturbing the electron beam, the mount must have a low magnetic susceptibility. In addition, the mount needs to be reasonably thermally conductive in order to allow heat to conduct out of the cavity. Aluminum satisfies these criteria with the added benefit that it is straightforward to machine.

Three piezoelectric actuators (PI P-883.10) which contact 50um thread pitch micrometer screws (Kozak Micro Adjusters TSBM5-05-10/7) are attached to one side of the flexure and press against the other side, allowing the second mirror to be tipped, tilted, and axially translated relative to the first mirror. The micrometer screws are used for coarse alignment when the distance to concentricity is greater than approximately 6um. This allows the cavity to be assembled with the mirrors several hundred microns from concentricity which accommodates the dimensional tolerances of the cavity mount and mirrors. However, when the cavity is aligned to near the concentric condition the orientation of the cavity mode becomes extremely sensitive to changes in the relative alignment of the two cavity mirrors, and fine alignment using the micrometer screws becomes difficult. At this point, the piezos (travel range of 6.5um) are used to complete the fine alignment to just 3.7um from concentricity.

As the cavity warms up during operation, the piezoelectric actuators are used to actively stabilize the position of the cavity mode on the mirrors. To avoid thermal stress on the mirrors from the differential thermal expansion of the mirrors and the aluminum mount, the mirrors are retained with flexure springs. The cavity design is shown in Figure 1(a).

## Mirrors

The limits on the cavity size constrain the maximum size of the mirrors. The outer diameter of the mirrors is 7.75mm. The radius of curvature on the concave (reflective) side of the mirrors is 10mm, making the distance between the concave surfaces of the mirrors when installed in the cavity 20mm minus the distance to concentricity. To minimize the length of the cavity mount, no additional coupling optics are used. The convex side of the mirrors has a radius of curvature of 5mm. A center thickness of 5.6mm allows the mirror to take a collimated laser beam incident on the convex surface and focus it at the center of the radius of curvature of the concave surface. The mirror mechanical information is shown in Figure 1(b). With this design, we have achieved a cavity coupling efficiency of approximately 90%, though due to drifting of various optical components, the cavity coupling is often lower. We tolerate operation of the cavity with coupling efficiencies as low as 65%.

For handling the high circulating power of the cavity, the mirrors must have minimal scattering. To achieve this, the mirrors were first polished to 20-10 scratch-dig surface quality on both the concave and convex sides, then superpolished on the concave side by Perkins Precision Developments. Final superpolishing of the concave side was done by Coastline Optics. After the superpolish, the mirrors have 1.0A RMS microroughness and 0-0 scratch-dig in the 3mm clear aperture. The surface figure deviates from spherical by less than 31.6nm over the central 6mm of the concave surface. The outer diameter of the mirrors is also polished (as opposed to ground) to reduce the risk of glass particles from the outer diameter of the mirrors contaminating the front surface.

The convex surface of the mirror has an ion-beam sputtered (IBS) dielectric anti-reflection coating with a reflectivity of <0.1%. The concave surface has a high-reflectivity quarter-wave optical thickness dielectric IBS coating with a transmissivity of 80ppm at a wavelength of 1064nm. Both coatings were done by FiveNine Optics.

To keep the scattering low and minimize mirror absorption, we take precautions to avoid contamination of the mirror surface after polishing and coating. The mirrors are only handled under ULPA filtered airflows, which have particle counts of <35/m$^3$ for particles >0.3um in diameter (conforming to ISO 3 or cleaner per ISO 14644-1). The cavity is also designed to limit paths for dust to reach the mirror from outside of the aluminum mirror mount. With these precautions, each mirror has optical losses (scattering and absorption) of around 5ppm. This constitutes 1.4W of lost power for a cavity circulating power of 140kW.

Even with a state-of-the-art low absorption (<1 ppm) IBS coating, the power absorbed by the mirrors is still appreciable. As a result, the temperature of the cavity-facing surface of the mirrors changes with the cavity circulating power. The temperature change is non-uniform, which causes variable thermal expansion over the cavity-facing surface of the mirror. The radius of curvature of the mirror thus increases, bringing the cavity farther from concentricity and increasing the mode waist at the center, as shown in Figure 3(d). Thermal expansion is mitigated by using premium grade ultra-low expansion (ULE) Corning 7972 glass, which has over an order of magnitude lower coefficient of thermal expansion than fused silica. Since ULE has layers, leading to striation planes in its refractive index, our mirrors are constructed such that these striations are normal to the optical axis.

## Cascaded Feedback Control

To meet the bandwidth and dynamic range requirements for actively maintaining resonance between the laser and the cavity with unusually high-power, we use three actuators in cascading feedback loops, as shown in Figure 2. Optomechanical instability[21] in high-power cavities is addressed by using a high-speed feedback loop to damp mechanical oscillations. The required bandwidth for this loop could most easily be achieved by varying the laser frequency. On the timescale of hours, heating from light scattered or absorbed increases the temperature of the cavity, causing a gradual thermal expansion and thus changing the resonant frequency of the cavity. Tracking this change requires a large feedback range, approximately 50 picometers. Figure 2 provides an overview of the laser frequency feedback loops, as well as other elements of the optical system.

For high-speed feedback (and lowest dynamic range), as well as 5 MHz frequency modulation sidebands for generating a Pound-Drever-Hall error signal, we use a fiber coupled acousto-optic modulator (AOM, G&H T-M150-0.4C2G-3-F2P) driven by a high-bandwidth voltage-controlled oscillator (VCO, Mini-Circuits ZX95-200A+). By setting the center frequency of the VCO, the residual amplitude modulation from the sidebands can be minimized, and this has been found to be stable over several months. This feedback loop is delay-limited, so bandwidth is limited by phase accumulation before appreciable amplitude roll-off occurs. The main sources of delay in the fastest loop are ~200 ns from the fiber AOM, ~80 ns from the seed laser traveling through the fiber amplifier (Azurlight Systems ALS-IR-1064-50-A-SF), and ~70 ns from the VCO. To control the AOM, we use a proportional-integrator-integrator controller. At lower frequencies, the double integrator maximizes the loop gain in order to suppress cavity mechanical resonances. The analog implementation of the controller minimizes the delay. The AOM and its controller provide effective high frequency feedback up to approximately 400kHz, but they can only provide a frequency shift up to about 5MHz.

A piezoelectric actuator in the seed laser (NKT Photonics Koheras ADJUSTIK K822) provides the next level of laser frequency feedback, with a larger range and lower bandwidth. The piezo can change the laser frequency by approximately +/-50MHz and its bandwidth is limited by a resonance at about 35kHz. Therefore, we use an FPGA based-proportional-integral controller (programmed on a Red Pitaya Stemlab 125-14), which achieves a 1 MHz bandwidth—well beyond the limit of the piezo—limited by delays due to the input analog to digital converter and output digital to analog converter. The input to this controller is the output of the AOM feedback controller, so that the FPGA controller keeps the AOM near its center frequency. By using the FPGA controller, we are able to add more complex conditional behavior to this feedback loop, including cavity autolocking and transient disturbance detection. For autolocking, a dip in the transmitted power triggers the controller to switch from providing feedback to scanning the piezo for the cavity resonance. A peak in the transmitted power then triggers a switch back to providing feedback, and the cavity reliably re-locks for cavity powers up to 40kW. The FPGA can also monitor the state of its own feedback loop and the faster AOM loop to determine if the loops are properly tracking the cavity resonance (the feedback loops may be failing to follow the cavity resonance if there is a disturbance which is too large or too fast, e.g. if a mechanical shock is applied to the cavity). By determining when the laser frequency feedback loops are stable, it is possible to prevent other feedback loops from actuating on incorrect information. In particular, we stop the final level of laser frequency feedback when the faster loops are giving inaccurate readings.

The final level of laser frequency feedback with the lowest bandwidth and greatest dynamic range is provided by temperature control of the seed laser. By changing the temperature of the seed laser substrate, the laser wavelength can be varied by +/-300pm. However, the bandwidth of this actuator is quite slow, limited to a few Hertz. The temperature setpoint is set by a software proportional-integral-derivative controller using the laser piezo setpoint as the error signal. The role of this slowest feedback loop is primarily to compensate for the thermal expansion of the cavity as it warms up. The temperature setpoint usually drifts by approximately 30pm from the time the laser is first locked to the cavity to the time when it reaches a stable operating condition, and subsequently changes only a few picometers per hour. The combination of these three actuators provides fast feedback over a large range, maintaining resonance between the laser and the cavity over the extent of cavity powers.

### Gain Stabilization

For the feedback loops to function properly, the error signal sensitivity should remain constant, in order to avoid drifts in the loop gain or setpoint. The reflected intensity signal at the sideband frequency is detected by a photodiode (Rx PD in Figure 2) and is given by

$$I_{r,\Omega} = \frac{k_R|Q|^2 I_i \beta}{2}\left[f_r(\omega)f_r^*(\omega+\Omega)e^{-i\Omega t} + f_r^*(\omega)f_r(\omega+\Omega)e^{+i\Omega t} - f_r^*(\omega)f_r(\omega-\Omega)e^{-i\Omega t} - f_r(\omega)f_r^*(\omega-\Omega)e^{+i\Omega t}\right]$$

where $k_R$ is the portion of reflected light reflected to the photodiode, $|Q|^2$ is the cavity coupling coefficient, $I_i$ is the intensity incident to the cavity, $\beta$ is the sideband modulation depth, $\omega$ is laser frequency relative to the cavity resonant frequency, $\Omega$ is the sideband frequency, and $f_r(\omega) = R - \frac{T^2 R e^{i\omega/\Delta_{fsr}}}{1 - R^2 e^{i\omega/\Delta_{fsr}}}$ is the cavity reflection transfer function, with R and T being the mirror reflectance and transmittance, respectively. We assume that the reflectance and transmittance for both mirrors are the same, though not necessarily real. The PDH error signal is generated by mixing the voltage at the photodiode, proportional to $I_{r,\Omega}$, with a reference signal. Thus, the PDH error signal is proportional to $I_{r,\Omega}$. The DC intensity at the photodiode is given by

$$I_{r,DC} = k_R|R|^2 I_i\left[1 + \frac{\beta^2}{2}\right] - |Q|^2|R|^2 I_i + |Q|^2 I_i |f_r(\omega)|^2 + |Q|^2 I_i \frac{\beta^2}{4}\left[|f_r(\omega+\Omega)|^2 + |f_r(\omega-\Omega)|^2\right]$$

with the same parameters. Although we change the laser power incident to the cavity while the feedback loops are running, we can easily keep the DC intensity at the photodiode constant with another feedback loop. Since the DC intensity and amplitude of the PDH error signal are both proportional to $I_i$, the feedback loop properly compensates for changing the incident power.

Thermal effects add a confounding factor: the cavity coupling coefficient $|Q|^2$ varies with incident power. Since the cavity lock can not be engaged at its full required operating power, the incident power must be increased during operation. As shown in Figure 3(b), the cavity coupling varies from 0.5 to 0.8. Although the change is relatively gradual, it leads to the gain of the feedback loop varying by nearly a factor of two. To maximize the noise suppression of the feedback loop, the loop is operated near closed-loop instability, so this change in gain is unacceptable. The PDH error signal and DC voltage do not have the same dependence on the cavity coupling, so the feedback loop actuating on the DC intensity does not help. To properly compensate for this effect, we measure the intensity of the light transmitted through the cavity.

The intensity signal at the sideband frequency at a photodiode in transmission (Tx PD in Figure 2) is given by

$$I_{t,\Omega} = \frac{k_T \beta |Q|^2}{2} I_i \left[ f_t(\omega) f_t^*(\omega+\Omega) e^{-i\Omega t} + f_t^*(\omega) f_t(\omega+\Omega) e^{+i\Omega t} - f_t(\omega) f_t^*(\omega-\Omega) e^{+i\Omega t} - f_t^*(\omega) f_t(\omega-\Omega) e^{-i\Omega t} \right]$$

where $f_t(\omega) = \frac{T^2}{1 - R^2 e^{i\omega/\Delta_{fsr}}}$ is the cavity transmission transfer function and $k_T$ is the portion of the transmitted light reflected to the photodiode, with all other parameters being the same as for reflection. The DC voltage on the transmission photodiode is given by

$$I_{t,DC} = k_T |Q|^2 I_i |f(\omega)|^2 + \frac{|Q|^2 I_i \beta^2}{4} \left[ |f(\omega+\Omega)|^2 + |f(\omega-\Omega)|^2 \right]$$

with the same parameters. In transmission, the PDH error signal and DC voltage have the same dependence on cavity coupling coefficient. However, a feedback loop using the transmitted PDH error signal would have an additional time delay, reducing its bandwidth.

In order to directly measure the gain of the feedback loop, we inject a small 1.1kHz modulation signal on the lockpoint. The feedback loop responds by changing the frequency of the laser proportional to the slope of the reflected PDH error signal. By measuring the amplitude of the 1.1kHz component of the transmitted PDH error signal, the slope of the reflected PDH error signal can be deduced. We then feedback on the DC voltage setpoint of the reflected photodiode in order to keep the reflected PDH error signal slope constant. The gain is proportional to the slope, so this procedure stabilizes the gain. By keeping the slope constant, this procedure also maintains the frequency offset produced by the feedback loop setpoint.

### Overview of the Electron Microscope

As previously described in Schwartz, 2019[19], the cavity is inserted into a custom FEI Titan, modified to support phase plate studies. In phase plate mode, the Lorentz lens of the microscope magnifies the back focal plane of the objective lens such that the effective focal length is approximately 20mm. A pair of 25mm diameter vacuum ports on the microscope is provided, which allows us to install the cavity in this plane. The image deflectors are used to steer the electron beam in the magnified back focal plane and align the undiffracted electron beam to a laser standing wave antinode. An additional lens, known as the transfer lens, then relays the image down to the usual set of projection lenses. After the transfer lens, an additional set of deflectors is used to align the electron beam to the downstream projection system (serving the function traditionally performed by the image deflectors). Apart from the additional transfer optics, the rest of the electron optics are unmodified.

These modifications to the microscope do have consequences. The magnification of the back focal plane and addition of the transfer lens substantially increase the chromatic aberration constant of the microscope by a factor of ~2.7. This increase causes a more severe temporal coherence envelope function, limiting the resolution that can be achieved with a dataset of practical size to approximately 3.5 Å. Furthermore, using the deflectors to move the diffraction pattern in order to align it to the stationary cavity introduces substantial additional coma from the transfer lens. Both of these issues will be mitigated in the future, but they currently limit the performance of the microscope.

## RESULTS

## Cavity Performance

As shown in Figure 3(a), the cavity can operate at circulating powers in excess of 120kW for several hours (with the operating time being limited by need, rather than by the cavity). At the time of writing, the cavity had accumulated over 260 hours operating at powers above 100kW. The mirror loss has remained below 5ppm throughout cavity operation so far, even following an incident in which the vacuum was vented with the cavity locked at high-power during microscope sample insertion. This suggests that the mirror coating does not significantly degrade at high-power, nor does dust accumulate on the mirrors. Despite the thermally induced change in the mirror radius of curvature, the mode waist remains below 8.5um with the cavity locked at high-power. The cavity is also an efficient power amplifier: the cavity coupling efficiency is typically between 65% and 90%, varying as the cavity support warms up, as shown in Figure 3(b). The maximum achievable cavity coupling efficiency has also remained stable over time, implying that the optical elements remain in good condition. With the mirror transmission of 80ppm and loss less than 5ppm, the power amplification factor exceeds 7,000, allowing us to use an input power of about 15W, which is easily obtained with commercial fiber amplifiers.

## Images of Test Specimens

We used the laser phase plate to collect cryo-EM images of frozen hydrated thermoplasma acidophilum 20S proteasomes. This ~750 kilodalton protein complex has become one of the standard test specimens to use when developing new imaging methods. The expression and purification of thermoplasma acidophilum proteasome is described by Yu, 2020[22]. 2.5 µl of purified proteasome sample (~1.0 mg/ml) was applied to glow-discharged holey carbon grids (Quantifoil 300 mesh Cu R1.2/1.3). The grids were blotted by Whatman No. 1 filter paper and plunge-frozen in liquid ethane using a Mark IV Vitrobot (Thermo Fisher Scientific) with blotting times of 12s at room temperature and over 90% humidity.

Images were recorded with the aid of SerialEM[23], which required only limited modification to deal with any new features associated with using the laser phase plate. The phase plate cavity and cavity support arm temperatures are sufficiently stable during data collection that the electron beam and laser beam alignments need to be adjusted no more often than every 30 minutes. Thus, although the electron beam and laser beam alignment would need to be automatically adjusted to facilitate collecting larger data sets, manual tuning was sufficient for this data set.

A total electron dose of 50 e-/Å$^2$ was used for each image stack, with ~1 e-/ Å$^2$/frame. The dose rate was ~6 $e^-$/pixel/s. We used a K2 direct detection camera in counting (although not super-resolution) mode. The pixel size at the specimen was 0.96 Å. We used a defocus of 1.5 µm (underfocused), which is a larger defocus than is optimal for the phase plate, but which makes exact comparison with non-phase plate data possible.

Sample images, after frame alignment, are shown in Figure 4. We collected 249 micrographs, of which over 40% exhibited phase shifts between 60° and 120°. The fraction of images with phase shifts in the range of 80° to 100° degrees is expected to increase substantially after automated phase monitoring and alignment stabilization are implemented.

**Reconstruction**

Using the 20S proteasome images, we obtained a 3D reconstruction using the Relion 3.1 workflow and unless otherwise stated all processing took place within the Relion 3.1 package[24]. The electron micrograph movies were aligned using MOTIONCOR2[25] and CTF estimation, including phase shift, was performed using CTFFIND 4.1[26]. Visual inspection of the motion-corrected micrographs was used to select a better quality subset of 132 micrographs. Laplacian-of-Gaussian autopicking was used to select an initial set of particles. The particles were extracted from the micrographs with a pixel size of 1.88 Å and box size of 128 pixels. 2D classification was run on this initial particle set and the best class images were used as 2D targets for a subsequent step of template-based particle picking. This approach to particle picking gave us 16451 particles that were then extracted. 6000 particles from the best 2D classes were also used in an Ab initio stochastic gradient descent refinement with D7 symmetry and a 200 Å mask to generate an initial 3D model. 3D classification was done and the particles in the class with the highest resolution features were selected for subsequent processing. These particles were re-extracted from motion corrected micrographs with a pixel size of 1.46 Å and a box size of 240 pixels. 3D refinement was carried out and alignment-free 3D classification of refined particles was used to further purify the pool of good particles. Multiple rounds of CTF refinement[27] and Bayesian polishing[28] were done followed by 3D refinement, until the gold standard resolution plateaued. Alignment-free 3D classification was again used to further purify the data followed by additional rounds of CTF refinement and particle polishing, until no further increases in resolution were detected.

Finally, a total of 4789 particles from 64 images were used in the final reconstruction. The average per-particle phase shift was 76°, with 80% of the particles having a phase shift between 60° and 90°, as shown in Figure 5(a). As mentioned previously, the images were not taken with the microscope in a coma-free condition; a residual beam tilt of 5.5 mrad was found and compensated for. The B-factor was estimated to be 131 $Å^2$ from twice the slope of the fit shown in Figure 5(b)[29]. The overall resolution of the reconstruction was 3.8 Å, as determined by the gold-standard Fourier shell correlation method[29]. The Fourier shell correlation curve is shown in Figure 5(c). Views of the reconstruction are shown in Figure 6(a) and (b).

## DISCUSSION

We successfully implement and use a near-concentric high-power Fabry-Perot cavity as a tunable phase plate with a phase shift of up to 90° for a 300 keV transmission electron microscope. Through use of state of the art mirrors and advanced feedback loops, we are able to reach and maintain the highest continuous laser intensity ever recorded.

Due to the additional chromatic aberration incurred by our use of the transfer lens system, the resolution of our reconstruction does not yet match that achieved by others for the 20S proteasome. A reconstruction using the Volta phase plate, for example, was achieved with a resolution of 2.4 Å with a B-factor of 103 $Å^2$ [12], and a resolution of 2.8 Å was achieved with a B-factor of approximately 80 $Å^2$ without using a phase plate[30]. By reducing the electron energy spread in our microscope, we expect the results obtained with the laser phase plate to become comparable or better.

In addition to addressing the chromatic aberration issue, we also intend to add automated monitoring and adjustment of the alignment of the electron beam to the laser standing wave. By doing so, the variation in phase shift between images should be reduced, increasing the fraction of images with useful particles. With these improvements, laser phase-contrast electron microscopy is expected to become routine. Indeed, we expect the improved low-frequency contrast demonstrated here will allow for results that extend beyond what can be achieved with current cryo-EM techniques.

Outside of transmission electron microscopy, these developments in high-power Fabry-Perot cavities have potential for other applications. The high laser intensity at the focus of the cavity may be used as a dipole trap[31]. The high intensity may also allow for observations of nonlinear behavior of light in vacuum[32].

## ACKNOWLEDGEMENTS

We thank B. Buijsse, W. Carlisle, R. Danev, P. Dona, S. Goobie, P. Grob, B. Hendriksen, G. W. Long, J. Lopez, D. Mastonarde, T. Nakane, E. Nogales, A. Rohou, and S. Scheres for advice and assistance with various aspects of the project.

This work has been supported by the U.S. National Institutes of Health grant No. 5 R01 GM126011-02, the Gordon and Betty Moore Foundation grant number 9366, a collaborative Research and Development agreement with ThermoFisher Scientific, and the Bakar Fellows Program. J. J. A. is supported by the National Science Foundation Graduate Research Fellowship Program Grant No. DGE 1752814. S. L. C. was supported by the Howard Hughes Medical Institute Hanna H. Gray Fellows Program Grant No. GT11085, and O. S. by the Human Frontier Science Program postdoctoral fellowship LT000844/2016-C.

**Figure 1**

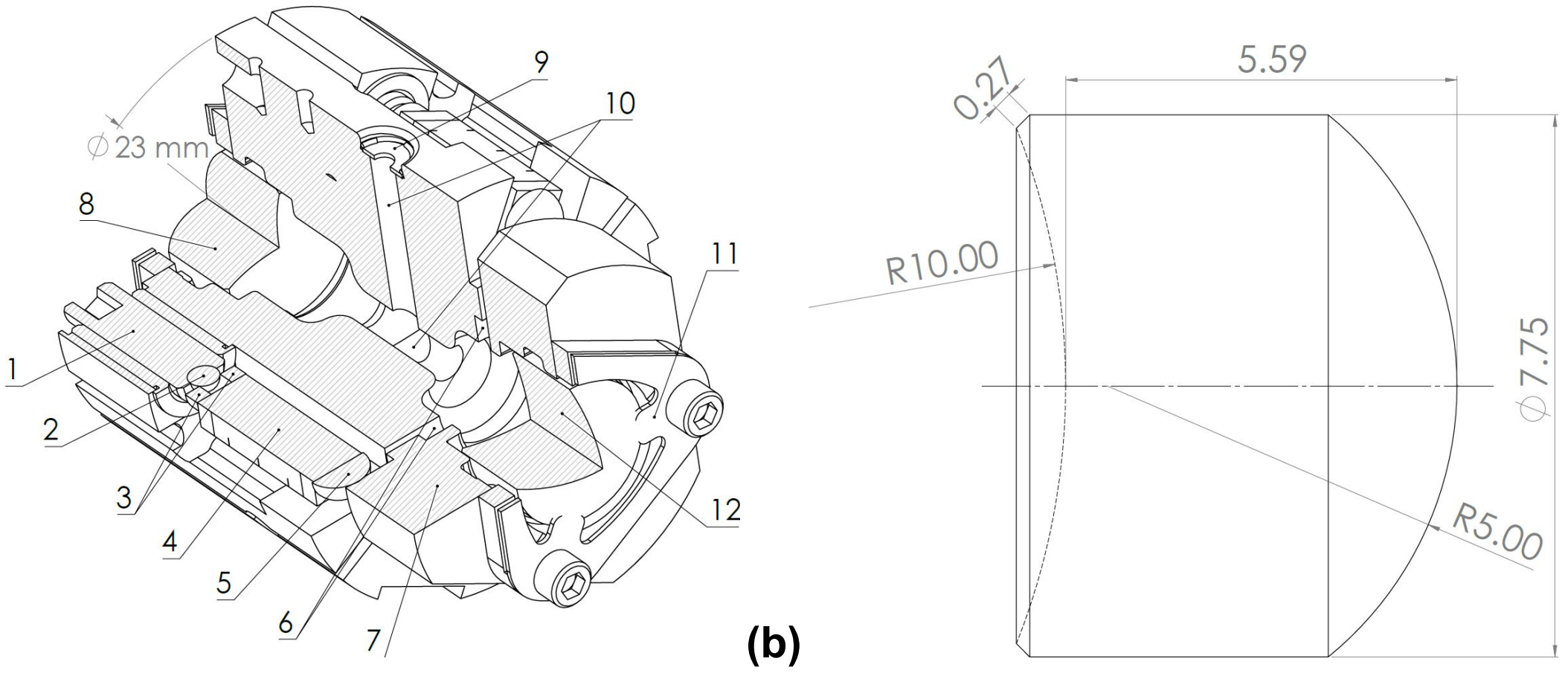

Cavity and mirror mechanical information. **(a)** Cross section view of the cavity with labelled features: 1) micrometer screw for coarse cavity alignment; 2) micrometer screw captive ball point (silicon carbide); 3) ring jewel bearing (sapphire) distributing the ball point pressure across the end of the 4) piezoelectric actuator; 5) hemispherical ball point (sapphire) to provide a pivoting contact point on the 7) moveable part of the cavity mount which moves via the 6) flexure spring cuts; 8) input mirror; 9) electron beam aperture (1.25 mm diameter, platinum-iridium) for the 10) electron beam channel; 11) flexure spring for axial mirror clamping; 12) output mirror. **(b)** Side view of the mirror geometry used for both the input and output mirrors. All dimensions are in mm.

**Figure 2**

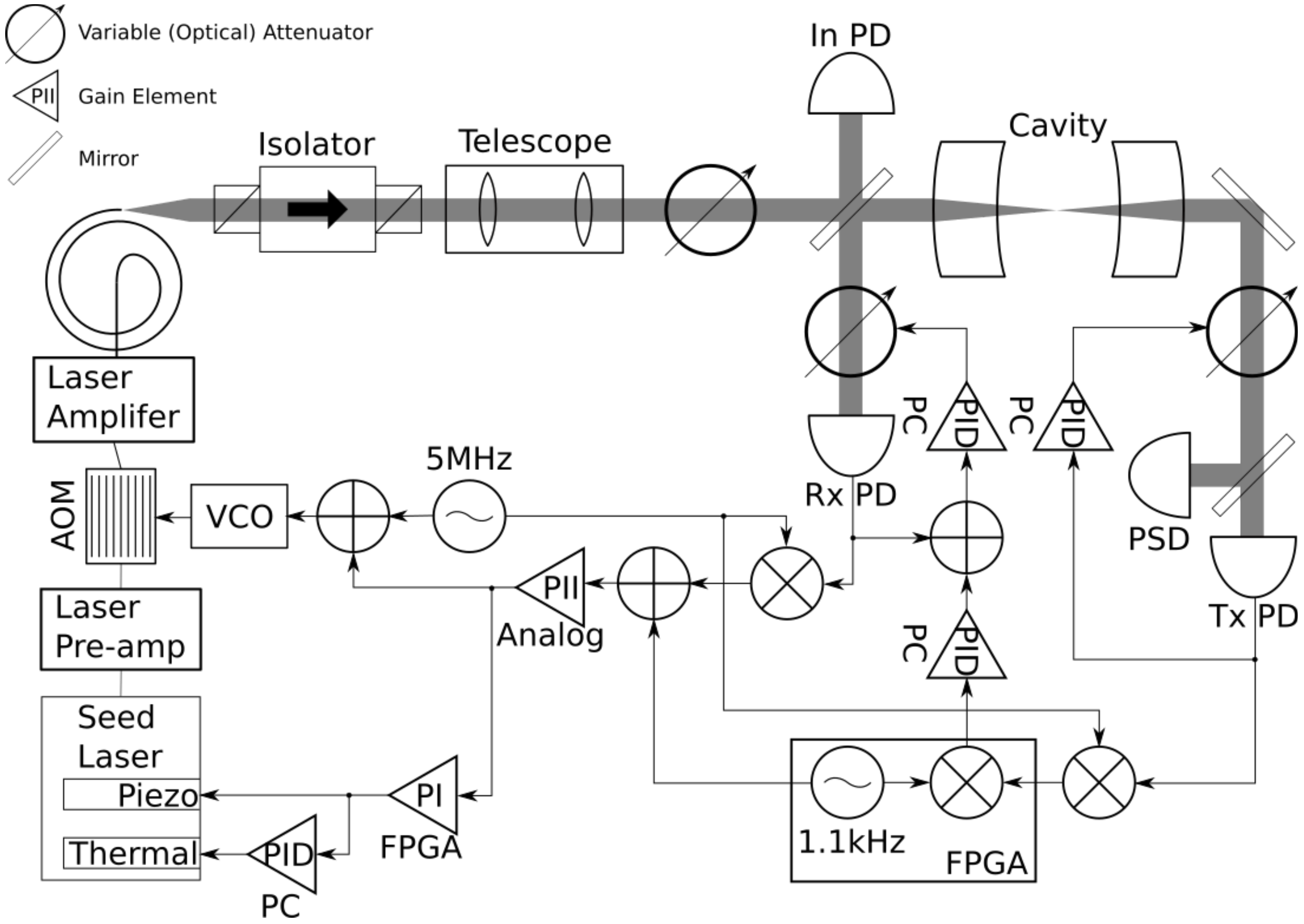

Schematic showing the optics and electronics required for the cavity. Light from the seed laser passes through a pre-amplifier (Azurlight Systems ALS-IR-106-0.05-A-SF) before the AOM in order to have sufficient intensity to seed the laser amplifier. The +1 diffracted order from the AOM is used. A telescope after the amplifier is used to match the size and divergence of the beam to the cavity mode. To avoid changing thermal lensing effects in the isolator and telescope, power incident to the cavity is varied using a variable attenuator consisting of a half-wave plate and (plate type) polarizing beamsplitter. A small fraction of the light incident to the cavity is picked off to monitor the incident intensity. A portion of the light reflected from the cavity is directed to the reflected light photodiode (Rx PD). A pair of polarizing beamsplitters and a half-wave plate is used as a variable attenuator to stabilize the power to the photodiode. Light transmitted through the cavity is similarly power stabilized before reaching a position sensitive detector (PSD) and the transmitted light photodiode (Tx PD). A 5MHz signal is sent to the AOM to generate sidebands for Pound-Drever-Hall stabilization of the laser frequency to the cavity frequency. To change the laser frequency, the AOM is controlled by an analog PII loop, the seed laser piezoelectric element is controlled by an FPGA-based PI loop, and the seed laser substrate heater is controlled by a PID loop running on a PC. Power to the photodiodes is stabilized using the DC voltage of the photodiode with a PID loop running on a computer. The reflected light DC setpoint is changed to stabilize the loop gain.

**Figure 3**

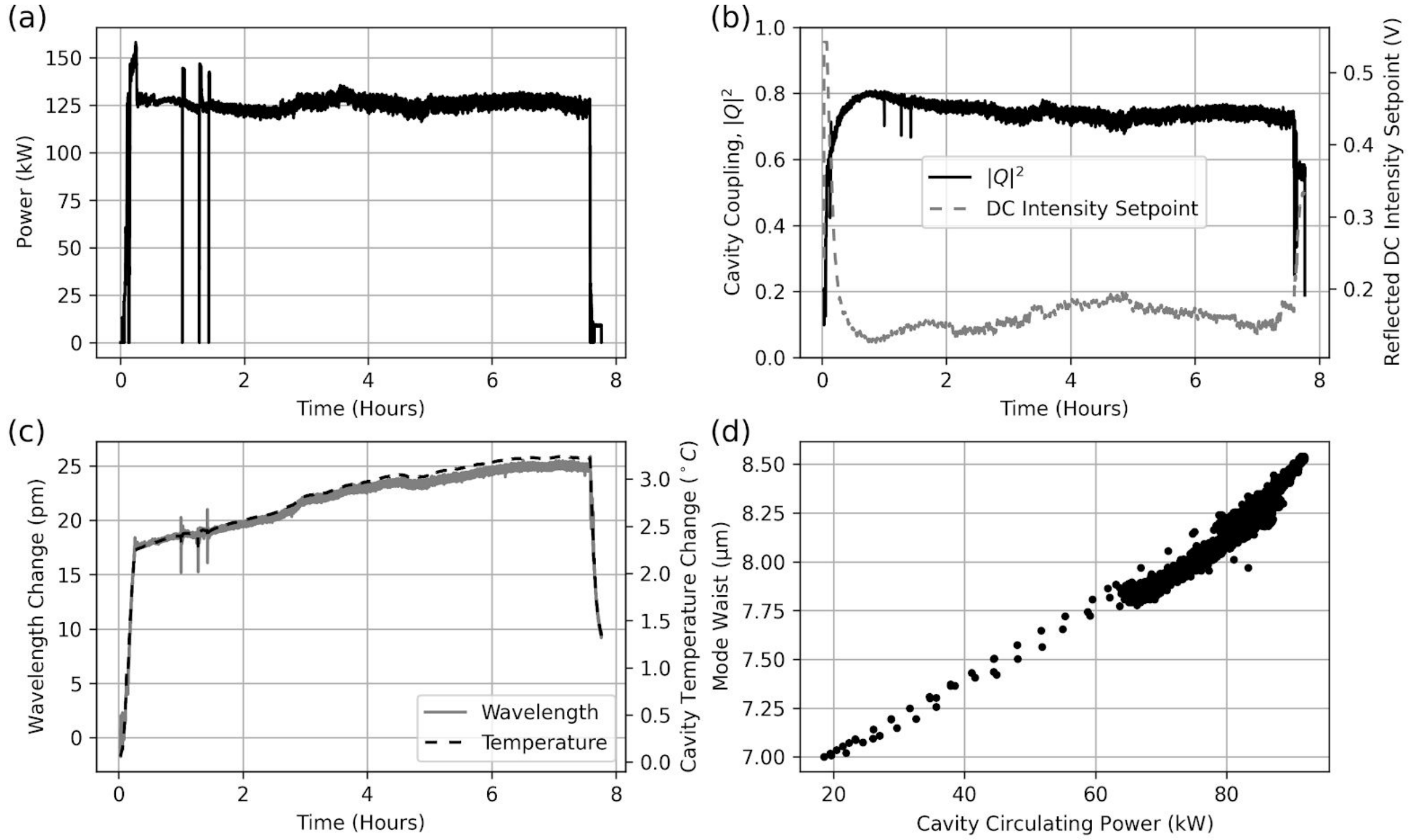

Cavity performance. **(a)** Cavity power over time for a typical data collection session. The short drops in cavity power are caused by cavity resonance being lost when microscope vacuum valves are actuated, which imparts a ~0.2g shock to the cavity. **(b)** Cavity coupling over time. The gain stabilization system varies the reflected DC intensity to keep the gain constant. **(c)** Change in cavity resonance wavelength over time. As the cavity warms up, thermal expansion changes the wavelength by approximately 25 picometers. **(d)** Change in mode waist as cavity power changes. The mode waist was measured using the technique described in Axelrod, 2020[18]. At higher powers, the increased mirror surface temperature increases the radius of curvature and causes the mode waist to increase.

**Figure 4**

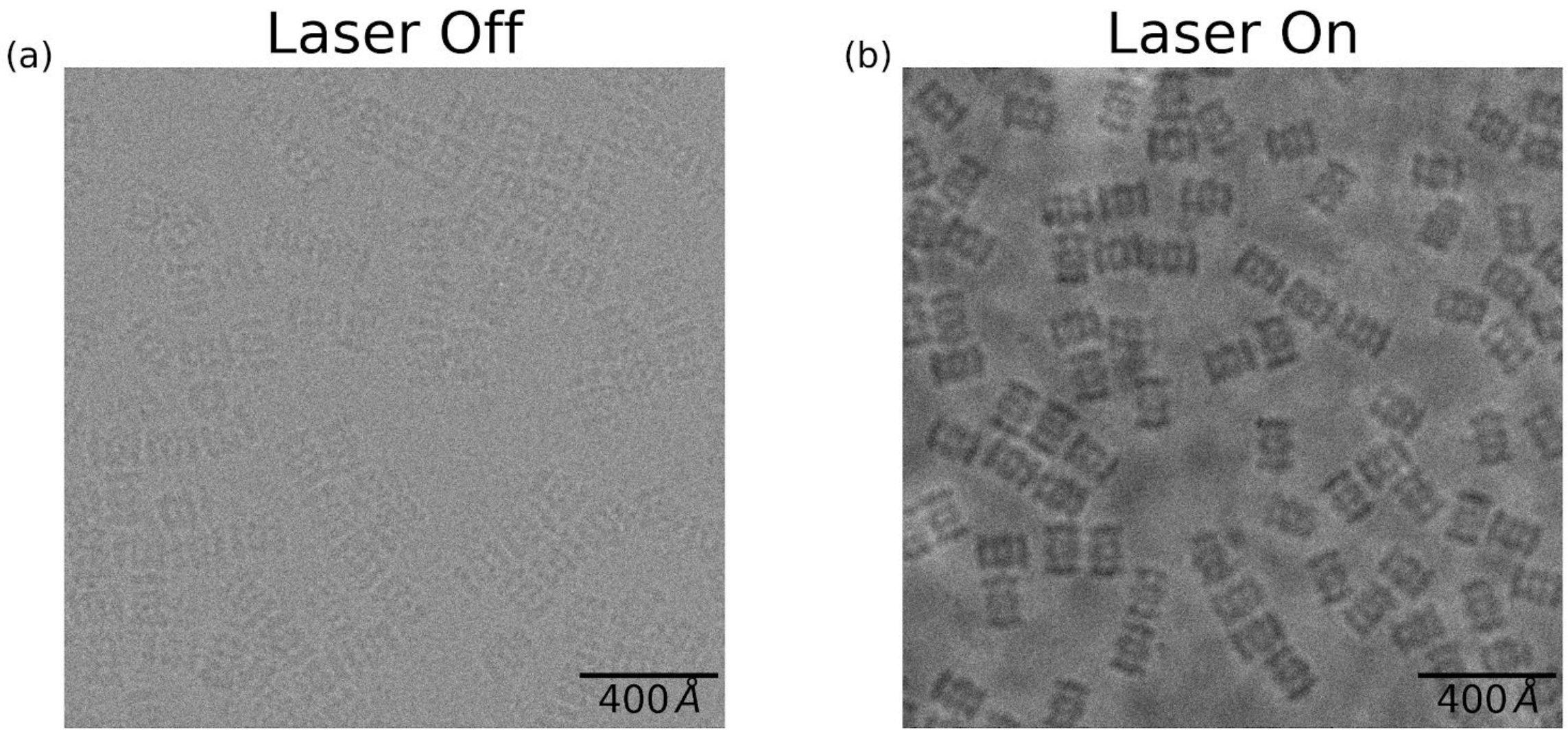

Images of 20S proteasome with and without the laser phase plate. Color scales are the same for both images in the figure. **(a)** Image of 20S proteasome sample taken without the laser phase plate at 1.5um defocus (underfocus). **(b)** Image of 20S proteasome sample taken with the laser phase plate at 1.5um defocus.

**Figure 5**

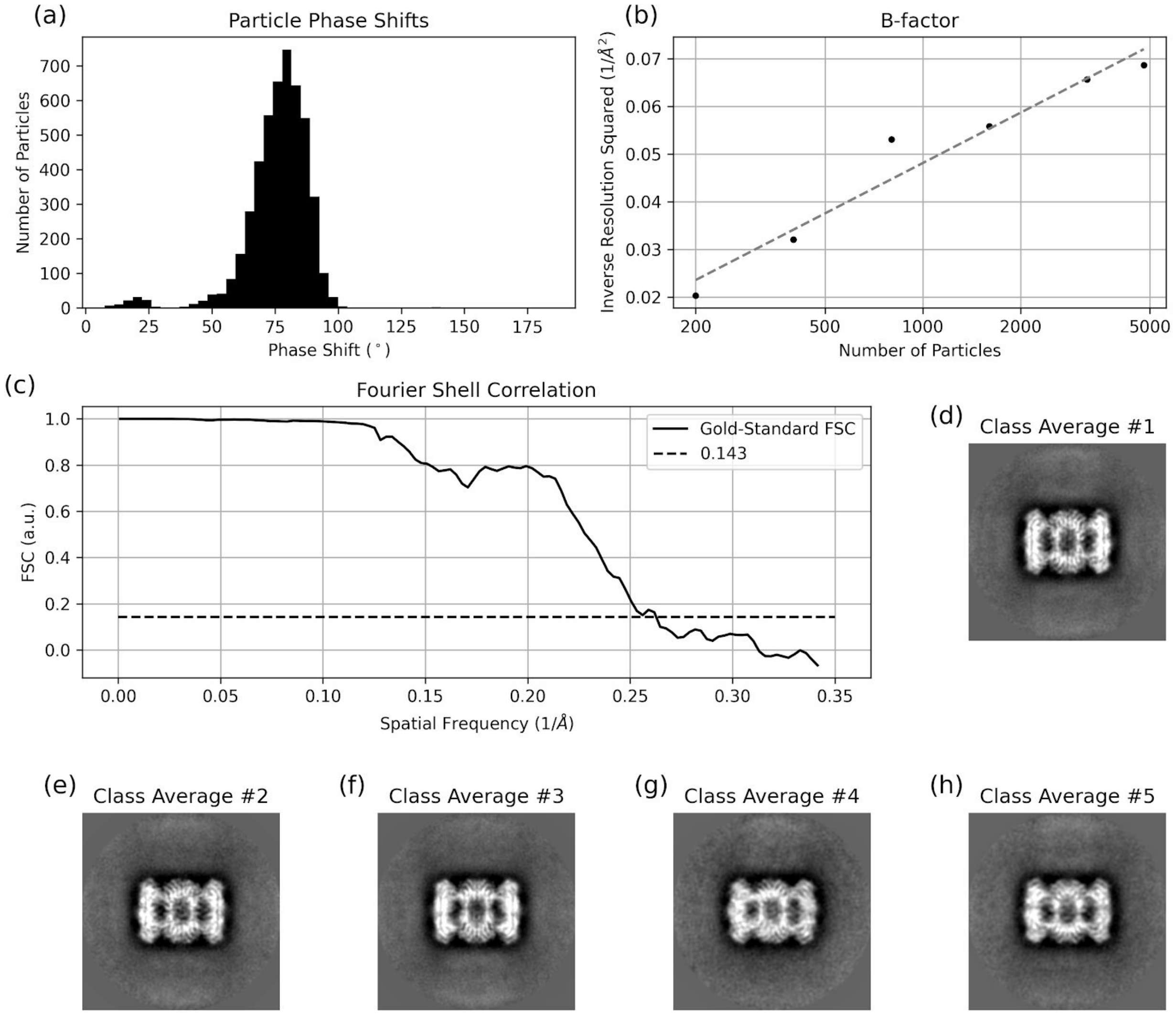

Results of the 20S proteasome reconstruction. **(a)** Per-particle phase shift histogram for the subset of the particles included in the final reconstruction. **(b)** B-factor curve. The B-factor for the reconstruction, calculated as two over the slope of the fit, was 131.3 Å$^2$. **(c)** Gold-standard Fourier shell correlation (FSC) curve. The FSC is generated by splitting the data into two sets, then correlating the two resulting reconstructions. The maximum resolution of the reconstruction, determined by where FSC falls below 0.143, was 3.8 Å. **(d-h)** Representative class averages for particles used in the reconstruction.

**Figure 6**

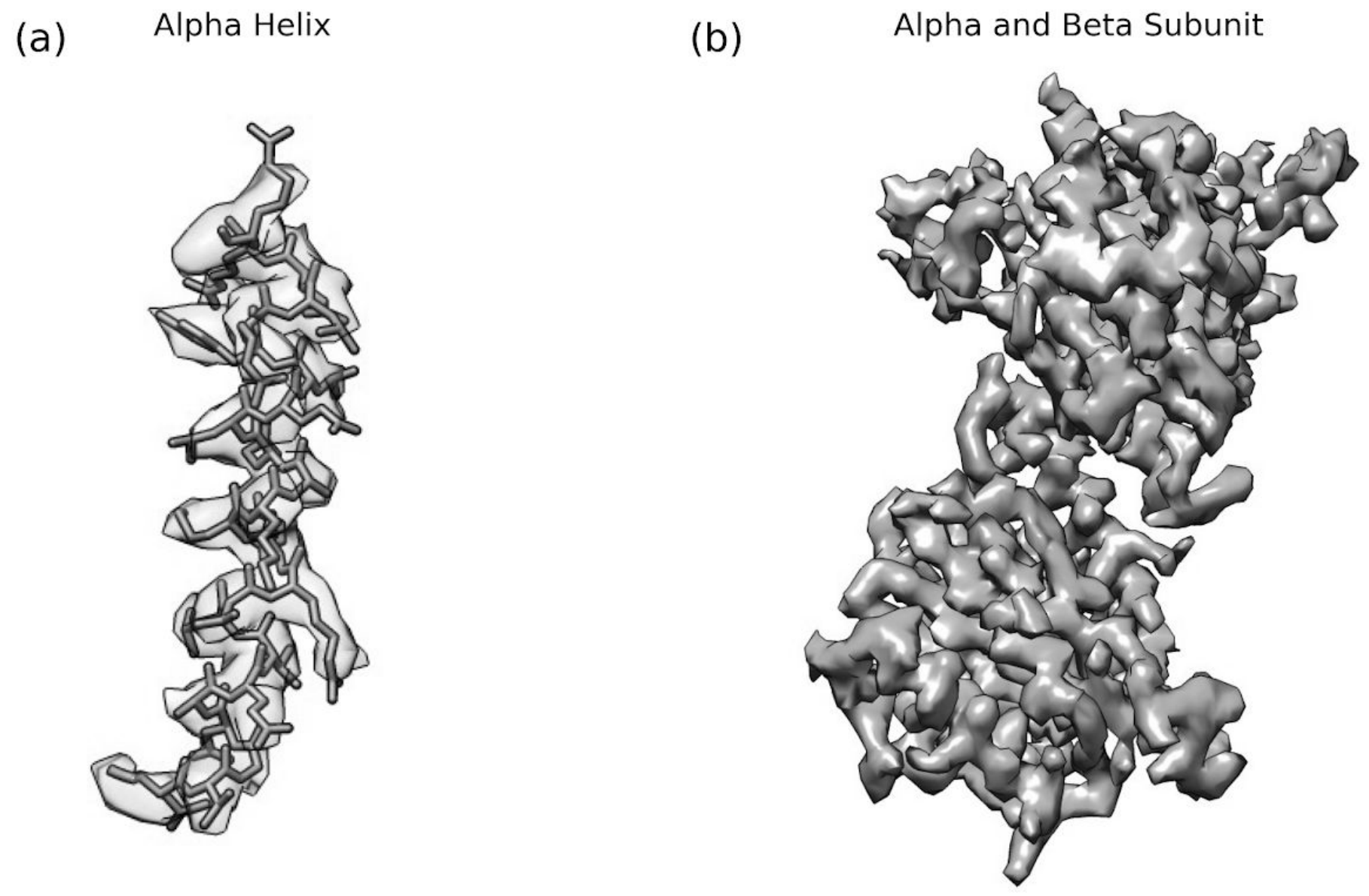

Results of the reconstruction. **(a)** An alpha helix showing residues 48-68 of the asymmetric unit of proteasome PDB 1PMA fit to our reconstruction's density map. **(b)** Alpha and beta subunit of the 20S proteasome.